\documentclass[epj]{svjour}
\usepackage{graphics}
\begin{document}
\newcommand*{\PKU}{School of Physics and State Key Laboratory of Nuclear Physics and
Technology, Peking University, Beijing 100871, China}
\title{Tri-meson-mixing of $\pi$-$\eta$-$\eta'$ and
$\rho$-$\omega$-$\phi$ in the light-cone quark model}

\author{Wen Qian \and Bo-Qiang Ma}

\institute{ {\PKU}} 
\date{Received: date / Revised version: date}
%
\abstract{
 The radiative transition
form factors of the pseudoscalar mesons {$\pi$, $\eta$, $\eta'$} and
the vector mesons {$\rho$, $\omega$, $\phi$} are restudied with
$\pi$-$\eta$-$\eta'$ and $\rho$-$\omega$-$\phi$ in tri-meson-mixing
pattern, which is described by tri-mixing matrices in the light-cone
constituent quark model. The experimental transition decay widths
are better reproduced with tri-meson-mixing than previous results in
a two-mixing-angle scenario of only two-meson $\eta$-$\eta'$ mixing
and $\omega$-$\phi$ mixing.
%
} 
\maketitle
\section{Introduction}
\label{intro} Light hadrons are well classified in SU(3) flavor
multiplets~\cite{Gell-Mann64Zweig64} based on
Gell-Mann~\cite{Gell-Mann61} and Ne'eman~\cite{Ne'eman61} theory.
The Gell-Mann-Okubo mass
formula~\cite{Gell-Mann61,Gell-Mann62Okubo62}, which exhibits the
mass relation of hadrons, has a simple linear form and a more
complicated quadratic form. In order to explain the deviation of the
meson masses from the Gell-Mann-Okubo mass formula, two-meson
$\omega$-$\phi$ mixing and $\eta$-$\eta'$ mixing were introduced
since the 1960s~\cite{Sakurai62Coleman64}. The mixing between mesons
helps to understand the SU(3) symmetry breaking.

The departure from the $\omega$-$\phi$ meson ideal mixing was %
studied~\cite{Okubo63} according to the meson mass spectrum, radiative decays %
and 2$\pi$, 3$\pi$ decay modes which are related to isospin conservation and %
Okubo-Zweig-Iizuka (OZI) rule violation. It is involved in the study of $\tau$ %
decays for a decade years~\cite{Castro96}. The mixing of $\eta$-$\eta'$ has also %
been studied based on pseudoscalar meson masses and their decays~\cite{Isgur76,Fritzsch77,Kawarabayashi81}. However, recently %
more analysis focus on their decay constants~\cite{Leutwyler98,Kaiser98,Feldmann98}, %
and a mixing scheme were promoted in which two mixing angles were introduced. %
The $\eta$-$\eta'$ mixing is also related to the study of $J/\psi$ decay~\cite{Okubo77}.

In fact, the isospin violating mixing of $\rho$-$\omega$ and
$\pi$-$\eta$ have been considered for a long time. $\rho$-$\omega$
mixing was first promoted by Glashow~\cite{Glashow61} based on
electromagnetic interaction. This mixing plays an important role in
the nuclear charge asymmetry study~\cite{Coon87}, and is related to
the CP violation analysis in hadronic B
decays~\cite{Enomoto,Gardner98,Guo07} and charmed hadron
decays~\cite{Guo00}. The off-shell behavior of $\rho$-$\omega$,
$\pi$-$\eta$ mixing was also studied~\cite{Maltman94,Piekarewicz93}.

Then it is natural to consider the tri-meson $\pi$-$\eta$-$\eta'$
mixing and $\rho$-$\omega$-$\phi$ mixing. Some efforts have been
devoted to them
respectively~\cite{Maltman94,Benayoun01,Benayoun08b,Benayoun08c,Gusbin81}.

Many different methods were used in above studies of meson mixing,
e.g., chiral perturbation theory, quantum chromodynamics(QCD) sum
rule, nonrelativistic and relativistic quark model, hidden local
symmetry (HLS) model, et al. The light-cone constituent quark
model~\cite{Lepage80,Brodsky82,Brodsky98} is also an important model
that has been used to study the mixing of
mesons~\cite{Choi97a,Cao98,Choi99,Xiao05,Huang07}. As far as
properties of mesons are studied at low energy scales,
nonperturbative QCD effects are important. The light-cone
constituent quark model is a convenient model for incorporating the
nonperturbative QCD effects effectively. With the Melosh-Wigner
rotation~\cite{Melosh74,Kondratyuk80,Ma93}, the SU(6) instant form
wave function can be transformed into the light-front form to
include relativistic effects. Taking the minimal quark-antiquark
Fock state description of photons and mesons, we can calculate their
transition form factors, decay widths and radii.

In the light-cone quark model, the mixing of pseudoscalar
$\eta$-$\eta'$ mesons has been studied using octet-singlet mixing
scheme~\cite{Donoghue85,Gilman87} and quark flavor
scheme~\cite{Feldmann99,Cao99}, and recently together with the
vector meson $\omega$-$\phi$ mixing, it is studied with a
two-mixing-angle scenario~\cite{Qian08}, which can reproduce data
better phenomenologically. When introducing two mixing angles, the
mixing matrices are non-unitary. Concerning the isospin symmetry
breaking, in this paper we restudy the transition form factors
between vector mesons and pseudoscalar mesons with tri-meson
$\rho$-$\omega$-$\phi$ mixing and $\pi$-$\eta$-$\eta'$ mixing in the
light-cone quark model, where the mixing is described by unitary
tri-mixing matrices~\cite{Benayoun01,Benayoun99}. This can also be
seen as an explanation of the  non-unitary mixing matrices  in the
two-mixing-angle scenario~\cite{Qian08}.

Since in this paper all the $I_z=0$ mesons are involved in mixing,
all the parameters have to be readjusted together without priority.
With the experimental data of the meson radii, decay widths as
constraints, we reset all the parameters in our model. New results
for the pion form factor between $Q^2=0.6$ and $2.45~\mathrm{GeV}^2$
at the Thomas Jefferson National Accelerator Facility (JLab) were
presented recently~\cite{Huber08}. We recalculate the $Q^2$ behavior
of the pseudo scalar meson form factors in our model using the new
parameters and compare them with the latest experimental data. The
 $Q^2$ behavior of the transition form
factors of other pseudoscalar and vector mesons are also
recalculated, and extrapolated to the time-like region and  compared
with experimental
data~\cite{Bebek78,Behrend91,Gronberg98,Acciarri98}.

This paper is organized as following. In Sec.~\ref{sec:level2}, we
explain the tri-meson-mixing of $\pi$-$\eta$-$\eta'$ and
$\rho$-$\omega$-$\phi$ respectively. In Sec.~\ref{sec:level3}, the
definitions of decay constants, form factors and radiative
transition form factors and their calculation in the light-cone
constituent quark model are briefly reviewed. In
Sec.~\ref{sec:level4}, we reset the parameters of our model by using
the experimental data as constraints. The $Q^2$ behaviors of form
factors are compared with experimental data.

\section{\label{sec:level2}Tri-meson-mixing of $\rho$-$\omega$-$\phi$ and $\pi$-$\eta$-$\eta'$}
The $\eta$-$\eta'$ and $\omega$-$\phi$ mixing were studied based on
different theories. Different values of mixing angles were
determined by different ways in literatures: the pseudoscalar meson
$\eta$-$\eta'$ mixing angle is around $20^\circ$ and the vector
meson $\omega$-$\phi$ mixing is around $28^\circ\sim 44^\circ$,based
on the Gell-Mann-Okubo mass formula, two-photon decay widths,
radiative and strong decay widths and so on~\cite{Feldmann00}.

Considering the discussion of pseudoscalar meson mixing  in the
chiral perturbation theory~\cite{Kaiser98,Leutwyler98,Feldmann99}
and PCAC formulas of flavour singlet current~\cite{Escribano05}, the
following mixing form
\begin{eqnarray}
\left(\begin{array}{c}
    |\eta\rangle \\
    |\eta'\rangle
    \end{array}\right)
&=&\left(\begin{array}{cc}
        \cos \theta^S_{qs} & -\sin \theta^S_{qs} \\
        \sin \theta^S_{qs} & \cos \theta^S_{qs} \\
    \end{array}\right)
    \left(\begin{array}{c}
        |\eta_q\rangle \\
        |\eta_s\rangle \\
    \end{array}\right),\\
\left(\begin{array}{cc}
    f_\eta^q & f_\eta^s \\
    f_{\eta'}^q & f_{\eta'}^s
    \end{array}\right)
&=&\left(\begin{array}{cc}
        \cos \theta^S_{qs} & -\sin \theta^S_{qs} \\
        \sin \theta^S_{qs} & \cos \theta^S_{qs} \\
    \end{array}\right)
    \left(\begin{array}{cc}
        f_q & 0 \\
        0 & f_s \\
    \end{array}\right),
\end{eqnarray}
with only one mixing angle is not enough to describe meson mixing.

  So in the light-cone constituent quark model
phenomenology studying, the meson mixing of $\eta$-$\eta'$ and
$\omega$-$\phi$ with two mixing angles are expressed as following
forms~\cite{Qian08}:
\begin{eqnarray}
\left(\begin{array}{c}
    |\eta\rangle \\
    |\eta'\rangle
    \end{array}\right)
&=&\left(\begin{array}{cc}
        \cos \theta^S_{q} & -\sin \theta^S_{s} \\
        \sin \theta^S_{q} & \cos \theta^S_{s} \\
    \end{array}\right)
    \left(\begin{array}{c}
        |\eta_q\rangle \\
        |\eta_s\rangle \\
    \end{array}\right),\\
\left(\begin{array}{c}
    |\phi\rangle \\
    |\omega\rangle
    \end{array}\right)
&=&\left(\begin{array}{cc}
        \cos \theta^V_{q} & -\sin \theta^V_{s} \\
        \sin \theta^V_{q} & \cos \theta^V_{s} \\
    \end{array}\right)
    \left(\begin{array}{c}
        |\omega_q\rangle \\
        |\omega_s\rangle \\
    \end{array}\right),
\end{eqnarray}
where, the quark-flavour basis mixing scheme is denoted by the
subscripts $qs$,  with quark flavor bases (for $\psi=\eta$ or
$\omega$)~\cite{Xiao05,Cao99}
\begin{eqnarray}
|\psi_q\rangle &=&
\frac{1}{\sqrt{2}}(u\bar{u}+d\bar{d})~\varphi^q(x,\mathbf{k}_\perp),\\
|\psi_s\rangle &=& s\bar{s}~ \varphi^s(x,\mathbf{k}_\perp).
\end{eqnarray}
In the octet-singlet scheme (denoted as
08)~\cite{Donoghue85,Gilman87}, just replace
$\{\theta_{qs}^{S,V},\eta_{q,s},\omega_{q,s}\}$ in above formula
with $\{\theta_{08}^{S,V}$, $\eta_{0,8}$, $\omega_{0,8}\}$. Here,
the flavor SU(3) octet basis and the singlet basis are (for
$\psi=\eta$ or $ \omega$).
\begin{eqnarray}
|\psi_8\rangle &=&
\frac{1}{\sqrt{6}}(u\bar{u}+d\bar{d})~\varphi_8^q(x,\mathbf{k}_\perp)
                -\frac{2}{\sqrt{6}}s\bar{s}~\varphi_8^s(x,\mathbf{k}_\perp), \\
|\psi_0\rangle &=&
\frac{1}{\sqrt{3}}(u\bar{u}+d\bar{d})~\varphi_0^q(x,\mathbf{k}_\perp)
                +\frac{1}{\sqrt{3}}s\bar{s}~\varphi_0^s(x,\mathbf{k}_\perp);
\end{eqnarray}
$\theta^{S,V}_{08}$ and $\theta^{S,V}_{qs}$ are respectively
pseudoscalar and vector meson mixing angles in the octet-singlet
scheme and the quark flavor scheme.

Meanwhile, the tri-mixing of pseudoscalar mesons
$\pi$-$\eta$-$\eta'$ and vector mesons $\rho$-$\omega$-$\phi$ have
been discussed respectively with different forms and
parametrization~\cite{Gusbin81,Maltman94,Benayoun01,Benayoun08b,Benayoun08c}.
Confined in a unitary mixing of the ideal basis, in this paper we
accept the tri-mixing matrix~\cite{Benayoun01} for describing the
meson mixing, as
\begin{eqnarray}
\left(
  \begin{array}{c}
  \omega \\
  \rho \\
  \phi\\
  \end{array}
\right) = MV \left(
  \begin{array}{c}
  \omega_I \\
  \rho_I \\
  \phi_I \\
  \end{array}
\right), \left(
  \begin{array}{c}
  \pi \\
  \eta \\
  \eta'\\
  \end{array}
\right) = MS \left(
  \begin{array}{c}
  \pi_I \\
  \eta_q \\
  \eta_s \\
  \end{array}
\right),
\end{eqnarray}
where,
\begin{eqnarray}
\begin{array}{lll}
\rho_I &=& \frac{1}{\sqrt{2}}(u\bar{u}-d\bar{d})\varphi_{\rho_I}, \\
\omega_I &=& \frac{1}{\sqrt{2}}(u\bar{u}+d\bar{d})\varphi_{\omega_I}, \\
\phi_I &=& -s\bar{s}\varphi_{\phi_I},
\end{array}
\begin{array}{lll}
\pi_I &=& \frac{1}{\sqrt{2}}(u\bar{u}-d\bar{d})\varphi_{\pi_I}, \\
\eta_q &=& \frac{1}{\sqrt{2}}(u\bar{u}+d\bar{d})\varphi_{\eta_q}, \\
\eta_s &=& s\bar{s}\varphi_{\eta_s},
\end{array}
\end{eqnarray}

\newsavebox{\triMix}
\begin{lrbox}{\triMix}
$MV = \left(
  \begin{array}{ccc}
    \cos\delta_V \cos\beta_V & -\sin\delta_V\cos\beta_V  & \sin\beta_V \\
    \sin\delta_V\cos\gamma_V + \cos\delta_V\sin\beta_V\sin\gamma_V & \cos\delta_V\cos\gamma_V-\sin\delta_V\sin\beta_V\sin\gamma_V & -\cos\beta_V\sin\gamma_V \\
    \sin\delta_V\sin\gamma_V - \cos\delta_V\sin\beta_V\cos\gamma_V & \cos\delta_V\sin\gamma_V+\sin\delta_V\sin\beta_V\cos\gamma_V & \cos\beta_V\cos\gamma_V \\
  \end{array}
\right).$
\end{lrbox}

\begin{equation}
\scalebox{0.6}[1]{\usebox{\triMix}}
\end{equation}

$MV$ is the vector meson tri-mixing matrix. $MS$ is the pseudoscalar
meson tri-mixing matrix and it has the same structure of $MV$ with
$\{\delta_V, \beta_V, \gamma_V\} $ replaced by
$\{\delta_S,\beta_S,\gamma_S\}$, that is, both the pseudoscalar
mixing and vector mixing are described by three mixing angles. Since
we do not discuss the CP violation and complex phases, the mixing
angles are all real in this paper. The wave function $
\varphi_{i}=A_i \exp[-\frac{m_u^2+\mathbf{k}_\perp^2} {8\beta_i^2
x(1-x)}]$, for $i=\omega_I, \rho_I,\pi_I,\eta_q$; $\varphi_{i}=A_i
\exp[-\frac{m_s^2+\mathbf{k}_\perp^2} {8\beta_i^2 x(1-x)}]$, for
$i=\phi_I,\eta_s$, where we adopt the Brodsky-Huang-Lepage (BHL)
prescription~\cite{Brodsky82,Huang94}.

Thus, the decay constants and transition form factors can be
calculated by the following tri-mixing formulas,
\begin{equation}
\left(
\begin{array}{l}
 F_{\pi\rightarrow\gamma\gamma^*} (Q^2) \\
 F_{\eta\rightarrow\gamma\gamma^*} (Q^2)  \\
 F_{\eta'\rightarrow\gamma\gamma^*}(Q^2)
\end{array}
\right) = MS \left(
\begin{array}{l}
 F_{\pi_I\rightarrow\gamma\gamma^*}(Q^2)  \\
 F_{\eta_q\rightarrow\gamma\gamma^*}(Q^2)  \\
 F_{\eta_s\rightarrow\gamma\gamma^*}(Q^2)
\end{array}
\right),
\end{equation}
\begin{equation}
\left(
\begin{array}{l}
 f_\omega  \\
 f_\rho  \\
 f_\phi
\end{array}
\right) = MV \left(
\begin{array}{l}
 f_{\omega_I} \\
 f_{\rho_I} \\
 f_{\phi_I}
\end{array}
\right),
\end{equation}
\begin{equation}
\left(
\begin{array}{l}
 F_{\omega \rightarrow\pi \gamma^* } (Q^2)\\
 F_{\omega \rightarrow\eta \gamma^* } (Q^2)\\
 F_{\eta' \rightarrow\omega \gamma^* }(Q^2) \\
 F_{\rho \rightarrow\pi \gamma^* }(Q^2) \\
 F_{\rho \rightarrow\eta \gamma^* } (Q^2)\\
 F_{\eta' \rightarrow\rho \gamma^* } (Q^2)\\
 F_{\phi \rightarrow\pi \gamma^* }(Q^2)\\
 F_{\phi \rightarrow\eta \gamma^* }(Q^2)\\
 F_{\phi \rightarrow\eta' \gamma^* }(Q^2)
\end{array}
\right) = ( MV\otimes MS )
 \left(
\begin{array}{l}
 F_{\omega_I\rightarrow\pi_I \gamma^* }(Q^2) \\
 F_{\omega_I\rightarrow\eta_q \gamma^* }(Q^2)\\
 0\\
 F_{\rho_I\rightarrow\pi_I \gamma^* } (Q^2)\\
 F_{\rho_I\rightarrow\eta_I \gamma^* }(Q^2) \\
 0\\
  0\\
  0 \\
  F_{\phi_I\rightarrow\eta_I \gamma^* }(Q^2)
\end{array}
\right).
\end{equation}

\section{\label{sec:level3}Decay constants and radiative transition form factors in the light-cone quark model}
In the $P\rightarrow\mu\nu$ process, the decay constant $f_P$ of the
charged pseudoscalar meson $P$ is defined as,
\begin{eqnarray}
\langle 0 |J^\mu|P(p)\rangle =i\sqrt{2}f_P p^\mu ,
\end{eqnarray}
where, $J^\mu$ is the axial-vector part of the charged weak current.
The charged form factor $F_P(Q^2)$ of a pseudoscalar meson $P$ is
\begin{eqnarray}
\langle
P(p')|J^\mu|P(p)\rangle\delta^3(\mathbf{p}-\mathbf{p}'+\mathbf{q}) =
F_P(Q^2)(p+p')^\mu ,
\end{eqnarray}
where, $J^\mu=e\bar{\psi}\gamma^\mu\psi$, $Q^2=-q^2$; and the
electromagnetic radius of the pseudoscalar meson is
\begin{eqnarray}
\langle r^2_P \rangle = -6 \frac{\partial F_P(Q^2)}{\partial
Q^2}|_{Q^2=0}.
\end{eqnarray}
The transition form factor of $P\rightarrow\gamma\gamma^*$ is
defined by~\cite{Jaus91,Choi97}
\begin{eqnarray}
&&\langle \gamma(p-q)|J^\mu |P(p,\lambda)\rangle \nonumber\\
&=& ie^2F_{P\rightarrow \gamma\gamma^*}(Q^2)
        \varepsilon^{\mu\nu\rho\sigma}p_\nu\epsilon_\rho(p-q,\lambda)
        q_\sigma,
\end{eqnarray}
so the decay width of $P\rightarrow\gamma\gamma$ is
\begin{eqnarray}
\Gamma(P\rightarrow\gamma\gamma) = \frac{1}{4}\pi\alpha^2 M_P^3
|F_{P\rightarrow\gamma\gamma^*}(0)|^2.
\end{eqnarray}

In the $V\rightarrow e^+e^-$ process, the decay constant of the
vector meson $V$ is defined by,
\begin{eqnarray}
\langle 0| j_\mu |V(p,S_z)\rangle=M_V f_V \epsilon_\mu(S_z).
\end{eqnarray}
Then, the decay width of $V\rightarrow e^+e^-$ is
\begin{eqnarray}
\Gamma(V\rightarrow e^+ e^-)=\frac{4\pi\alpha^2f_V^2}{3M_V}.
\end{eqnarray}

The radiative transition form factor between a pseudoscalar meson
$P$ and a vector meson $V$ is  defined by~\cite{Choi97}
\begin{eqnarray}
&& \langle P(p')|J^\mu |V(p,\lambda)\rangle  \nonumber\\ %
& =& ieF_{V\rightarrow P\gamma}(Q^2)
        \varepsilon^{\mu\nu\rho\sigma}\epsilon_\nu(p,\lambda) p'_\rho
        p_\sigma,
\end{eqnarray}
where, $\epsilon(p,\lambda)$ is the polarization vector of the
vector meson. Thus the decay widths of $V\rightarrow P\gamma$ and
$P\rightarrow V\gamma$ are
\begin{eqnarray}
\Gamma_{V\rightarrow P\gamma}
    &=&\frac{\alpha}{3}
    \left|F_{V\rightarrow P\gamma^*}(0)\right|^2
    \left(\frac{M_V^2-M_P^2}{2 M_V}\right)^3, \\
\Gamma_{P\rightarrow V\gamma}
    &=&\alpha\left|F_{P\rightarrow V\gamma^*}(0)\right|^2
    \left(\frac{M_P^2-M_V^2}{2 M_P}\right)^3.
\end{eqnarray}

To calculate the above decay constants and transition form factors
of pseudoscalar and vector mesons, we use the light-cone quark model
with the lowest Fock state expansion as an approximation. The wave
function of a meson is simplified to be
\begin{eqnarray}
&& |M (P^+, \mathbf{P}_\perp, S_z) \rangle \nonumber\\
   &=&\sum_{\lambda_i}\int\prod_{i=1}^2 \frac{\mathrm{d} x_i \mathrm{d}^2
        \mathbf{k}_{\perp i}}{\sqrt{x_i}~16\pi^3}
        16\pi^3\delta(1-\sum_{i=1}^2 x_i)\delta^{(2)}(\sum_{i=1}^2 \mathbf{k}_{\perp
        i})\nonumber\\
   && \,\,\times\,\,    | n : x_i P^+, x_i \mathbf{P}_\perp+\mathbf{k}_{\perp i},
        \lambda_i \rangle
        \psi_{2/M}(x_i,\mathbf{k}_{\perp i},\lambda_i) \nonumber\\
   &=&  \int \frac{\mathrm{d} x \mathrm{d}^2
        \mathbf{k}_{\perp}}{\sqrt{x(1-x)}16\pi^3}
        \varphi(x,\mathbf{k}_{\perp})
        \chi_M^{S_z}(x,\mathbf{k}_{\perp},\lambda_1,\lambda_2),
\end{eqnarray}
where, the momentum-space wave function
$\varphi(x,\mathbf{k}_\perp)$ is described by the BHL
prescription~\cite{Brodsky82,Huang94}
\begin{eqnarray}
\varphi(x,\mathbf{k}_{\perp}) &=&
\varphi_{\mathrm{BHL}}(x,\mathbf{k}_{\perp})\nonumber\\
&=& A \exp
\left[-\frac{1}{8\beta^2}\left(\frac{m_1^2+\mathbf{k}_\perp^2}{x}
+\frac{m_2^2+\mathbf{k}_\perp^2}{1-x}\right)\right],
\end{eqnarray}
and $\chi_M^{S_z}(x,\mathbf{k}_{\perp},\lambda_1,\lambda_2)$ is the
spin wave function.

 Through the Melosh-Wigner rotation
\cite{Kondratyuk80,Melosh74,Ma93}
\begin{equation}
\left\{
\begin{array}{lll}
\chi_i^\uparrow(T) &=& w_i[(k_i^+ +m_i)\chi_i^\uparrow(F)-k_i^R
\chi_i^\downarrow(F)]\\
\chi_i^\downarrow(T)&=& w_i[(k_i^+ +m_i)\chi_i^\downarrow(F)+k_i^L
\chi_i^\uparrow(F)]
\end{array}
\right. ,
\end{equation}
where $w_i=1/\sqrt{2k_i^+ (k^0+m_i)}$, $k^{R,L}=k^1\pm k^2$,
$k^+=k^0+k^3=x \mathcal{M}$,
$\mathcal{M}=\sqrt{\frac{\mathbf{k_\perp}^2+m_1^2}{x}+\frac{\mathbf{k_\perp}^2+m_2^2}{1-x}}$,
or corresponding proper meson vertex~\cite{Choi97,Yu07},
\begin{equation}\label{wavefucntion}
\bar{u}(k_1,\lambda_1)\Gamma_{M}v(k_2,\lambda_2),
\end{equation}
with
\begin{eqnarray}
\Gamma_P &=& \frac{1}{\sqrt{2}
\sqrt{\mathcal{M}^2-(m_1-m_2)^2}}\gamma_5,\ \mathrm{for\
pseudoscalar\ mesons, } \nonumber\\
\Gamma_V &=&-\frac{1}{\sqrt{2} \sqrt{\mathcal{M}^2-(m_1-m_2)^2}}
 (\gamma^\mu-\frac{k_1^\mu-k_2^\mu}{\mathcal{M}+m_1+m_2})\nonumber\\
 &&\times\ \epsilon_\mu(P,S_z),\ \mathrm{for\ vector\ mesons,}
\end{eqnarray}
we get the light-cone spin wave function of mesons and calculate the
above decay constants and transition form factors. The formulas were
listed in the appendix of Ref.~\cite{Qian08}.

\section{\label{sec:level4}Numerical Calculation Results}
    In the light-cone constituent quark model, when only one mixing
angle is introduced to describe the pseudoscalar meson mixing, the
angle can actually be analytically determined by the
$Q^2\rightarrow\infty$ limit behavior of $F_{P\gamma\gamma^*}(Q^2)$
~\cite{Cao99,Xiao05}. However, when two or three mixing angles are
introduced, there is no more analytic solution. So in this paper,
the meson mixing angles and all the parameters in wavefunction are
determined phenomenologically by electroweak properties of the
mesons by fitting light-cone constituent quark model results to
experimental data.

  In this paper, all parameters
including $m_u$, $m_s$, wave function parameters of $\pi$, $K$,
$\eta_q$, $\eta_s$, $\rho_I$, $\omega_I$, $\phi_I$ are set together
by using the experimental data, which include electromagnetic decay
widths and radii. We assume that the wave function parameters of
$\pi^\pm$ are the same as those of $\pi_I$. Though the off-shell
behavior of $\pi$-$\eta$-$\eta'$ has been discussed in some papers,
we assume the mixing angles to be constants when $Q^2$ changing. The
$Q^2$ behavior of the radiative transition form factors are
recalculated using the reset parameters and compared with
experimental data. The input data are all from PDG08~\cite{PDG08},
as listed in Table~\ref{tab:table1}.

The $Q^2\rightarrow\infty$ limiting behavior of $Q^2
F_{P\rightarrow\gamma\gamma^*}$ is also considered as constraints
for setting the parameters~\cite{Lepage80,Cao99}:
\begin{eqnarray}
\lim_{Q^2\rightarrow\infty}Q^2F_{P\rightarrow\gamma\gamma^*}(Q^2) =
2c_P f_P = \frac{2 c_P^2}{4\pi^2 F_{P\rightarrow\gamma\gamma^*}(0)},
\end{eqnarray}
where,
\begin{eqnarray}
c_P =
(c_{\pi_I},c_{\eta_q},c_{\eta_s})=(1,\frac{5}{3},\frac{\sqrt{2}}{3}).
\end{eqnarray}

The results are shown in Table~\ref{tab:table1}, together with the
results of Ref.~\cite{Qian08}. As we can see, the reproduction of
the radiative decay widths are improved with tri-meson-mixing
pattern. In fact, since $\beta_S,\delta_S\sim 0$, the mixing matrix
$MS$ is similar to the mixing of $\eta$, $\eta'$, and in the fit
result $\gamma_S=39.40^\circ$ is compatible with the results we got
in the two-mixing angle scenario of the quark flavor scheme
$\theta_q^S=40.57^\circ$, $\theta_s^S=43.89^\circ$. The vector meson
mixing bases are a little different from the ones in the two-mixing
angle scenario of the quark flavor scheme, which is equivalent to a
$90^\circ$ rotation: $90^\circ-\theta_q^V=3.29^\circ$,
$90^\circ-\theta_s^V=-3.43^\circ$. The $\beta_V=-6.2^\circ$ got in
the tri-meson-mixing fit is also compatible with them.

 There were following form relations between
 the mixing angles introduced in Ref.~\cite{Feldmann99},
\begin{eqnarray}
\theta_8 = \phi - \arctan \frac{\sqrt{2}f_s}{f_q}\nonumber\\
\theta_0 = \phi - \arctan \frac{\sqrt{2}f_q}{f_s}.
\end{eqnarray}
When three mixing angles are introduced, relations are not so simple.
 Similar to derivation in Ref.~\cite{Feldmann99},
\begin{eqnarray}
\left(
  \begin{array}{ccc}
    f_\pi^{\pi} & f_\pi^q & f_\pi^s \\
    f_\eta^{\pi} & f_\eta^q & f_\eta^s \\
    f_{\eta'}^{\pi} & f_{\eta'}^q & f_{\eta'}^s \\
  \end{array}
\right) %
= MS (\delta_S, \beta_S, \gamma_S) %
\left(
  \begin{array}{ccc}
    f_\pi & 0 & 0 \\
    0 & f_q & 0 \\
    0 & 0 & f_s \\
  \end{array}
\right),
\end{eqnarray}

\begin{small}
\begin{eqnarray}
\left(
  \begin{array}{ccc}
    f_\pi^{\pi} & f_\pi^8 & f_\pi^0 \\
    f_\eta^{\pi} & f_\eta^8 & f_\eta^0 \\
    f_{\eta'}^{\pi} & f_{\eta'}^8 & f_{\eta'}^0 \\
  \end{array}
\right) %
&=& MS (\delta_S, \beta_S, \gamma_S)%
\left(
  \begin{array}{ccc}
    f_\pi & 0 & 0 \\
    0 & f_q & 0 \\
    0 & 0 & f_s \\
  \end{array}
\right)%
MS^\dagger (0, 0, \gamma_{\mathrm{ideal}}),\nonumber\\
\end{eqnarray}
\end{small}
where, $\gamma_{\mathrm{ideal}}=\arctan\sqrt{2}$.
$\{f_\pi^8, f_\pi^0, f_\eta^\pi, f_{\eta'}^\pi\}$ are very small, if they are ignored, %
two mixing schemes can be related by
\begin{small}
\begin{eqnarray}
\left(
  \begin{array}{ccc}
    f_\pi^{\pi} & f_\pi^8 & f_\pi^0 \\
    f_\eta^{\pi} & f_\eta^8 & f_\eta^0 \\
    f_{\eta'}^{\pi} & f_{\eta'}^8 & f_{\eta'}^0 \\
  \end{array}
\right) %
\sim \left(
  \begin{array}{ccc}
    f_\pi^{\pi} & 0 & 0 \\
    0 & f_\eta^8 & f_\eta^0 \\
    0 & f_{\eta'}^8 & f_{\eta'}^0 \\
  \end{array}
\right) %
= \left(
  \begin{array}{ccc}
    f_\pi^{\pi} & 0 & 0 \\
    0 & f_8\cos\theta_8^S & -f_0\sin\theta_0^S \\
    0 & f_8\sin\theta_8^S & f_0\cos\theta_0^S \\
  \end{array}
\right) , \nonumber\\%
\end{eqnarray}
\end{small}
that leads to
\begin{eqnarray}
\begin{array}{c}
  \tan\theta_8^S = %
    \frac{\cos\delta_S\tan\gamma_S+\sin\beta_S\sin\delta_S-\sqrt{2}\frac{f_s}{f_q}\cos\beta_S}
    {\cos\delta_S-\sin\beta_S\tan\gamma_S\sin\delta_S+\sqrt{2}\frac{f_s}{f_q}\cos\beta_S\tan\gamma_S},    \\
  \tan\theta_0^S = %
    \frac{\cos\beta_S \tan\gamma_S - \sqrt{2}\frac{f_q}{f_s}(\cos\delta_S-sin\beta_S\sin\delta_S\tan\gamma_S)} %
    {\cos\beta_S - \sqrt{2}\frac{f_q}{f_s}(\cos\delta_S\tan\gamma_S -
    \sin\beta_S\sin\delta_S)}.
\end{array}
\label{eq:AngleRel}
\end{eqnarray}
Here, the $f_\pi^\pi$, $f_q$, $f_s$ are respectively equal to
 $\frac{1}{4\pi^2 F_{\psi\gamma\gamma^*}(0)}$ (for $\psi=\pi_I,\eta_q,\eta_s$) .

With the parameters set in this paper, we have
\begin{eqnarray}
\begin{array}{c}
  f_\pi^\pi = 0.091 \ \mathrm{GeV}, \\
  f_q =  0.097 \ \mathrm{GeV},\\
  f_s = 0.194\ \mathrm{GeV}.
\end{array}
\end{eqnarray}
Putting the $\beta_S$, $\delta_S$, $\gamma_S$ shown in the fifth
column of Table~\ref{tab:table1}
 into Eqs.(\ref{eq:AngleRel}), one can get $\theta_8^S = -30.96^\circ$, and $\theta_0^S = 3.93^\circ$.
 These two values are close to the results we got when using the two-mixing angle scheme
 as shown in the third column of Table~\ref{tab:table1}.

\begin{table*}
\caption{ Experimental data~\cite{PDG08} for the decay constants and
decay widths of $\eta$, $\omega$, $\phi$ are compared with
theoretical values. The first column is the experimental data. The
second and third column are the theoretical results of two-mixing
angle scenario~\cite{Qian08}. The fourth column is the theoretical
result with tri-meson-mixing pattern. Parameters set in different
schemes are listed below.  } \label{tab:table1}
\begin{tabular}{ccccccc}
    & $F_{\mathrm{exp}}/f_{\mathrm{exp}}~(\mathrm{GeV})$
     & $\begin{array}{c}
        F_{\mathrm{th}}/f_{\mathrm{th}}~(\mathrm{GeV}) \\($two-angle$ \\ $08 scheme$)
         \end{array}$
     & $\begin{array}{c}
        F_{\mathrm{th}}/f_{\mathrm{th}}~(\mathrm{GeV}) \\($two-angle$ \\ $qs scheme$)
         \end{array}$
     &  $\begin{array}{c}
        F_{\mathrm{th}}/f_{\mathrm{th}}~(\mathrm{GeV}) \\($tri-meson-mixing$)
         \end{array}$\\
  \hline
    $f_{\pi^+}$  & $0.0922\pm0.0001$  &  $0.0922$ &  $0.0922$ & $0.0920$\\
    $\langle r_\pi^2 \rangle$ $\mathrm{fm^2}$   & $0.45\pm0.01$   &  $0.45$ &  $0.45$ & $0.45$\\
    $F_{\pi^0\rightarrow\gamma\gamma^*}(0)$ & $0.274\pm0.010$      &  $0.274$ &  $0.274$ & 0.279\\
    $f_{K^+}(K^+\rightarrow \mu\nu)$  & $0.1100\pm0.0006$   &  $0.1100$ &  $0.1100$ &  0.1106\\
    $\langle r_{K^+}^2 \rangle$ $\mathrm{fm^2}$   & $0.31\pm0.03$ &  $0.31$ &  $0.31$ & 0.31\\
    $\langle r_{K^0}^2 \rangle$ $\mathrm{fm^2}$ & $-0.077\pm0.010$    & $-0.077$ & $-0.077$ & -0.077\\
  $F_{\eta\rightarrow\gamma\gamma^*}(0)$  & $0.272\pm0.007$     & 0.272 & 0.259 & 0.277\\
  $F_{\eta'\rightarrow\gamma\gamma^*}(0)$ & $0.342\pm0.006$   & 0.342 & 0.317 & 0.334\\

    $f_{\rho}(\rho\rightarrow e^+ e^-)$  & $0.1564\pm0.0007$  &  0.1564 &  0.1564  & 0.1603 \\
  $f_\phi(\phi\rightarrow e^+ e^-)$  & $0.076\pm 0.012$  & 0.068 & 0.076 & 0.075 \\
  $f_\omega(\omega\rightarrow e^+ e^-)$ & $0.0459\pm0.0008$  & 0.0475 & 0.0456  & 0.04556 \\
  $F_{\rho^+\rightarrow\pi^+\gamma}(0)$ & $0.83\pm0.06$ & 0.83 & 0.83 & 0.84 \\
  $F_{\phi\rightarrow\pi\gamma^*}(0)$    & $0.133\pm0.003$   & 0.131 & 0.132  & 0.132 \\
  $F_{\omega\rightarrow\pi\gamma^*}(0)$    & $2.385\pm0.004 $  & 2.327 & 2.295  & 2.382\\

  $F_{\phi\rightarrow\eta\gamma^*}(0)$    & $-0.692\pm0.007 $   & $-0.581$ & $-0.662$  & -0.677 \\
  $F_{\phi\rightarrow\eta'\gamma^*}(0)$    & $0.712\pm0.01$     & 0.853 & 0.742  & 0.727 \\
  $F_{\omega\rightarrow\eta\gamma^*}(0)$    & $0.449\pm0.02$   & 0.453 & 0.457  & 0.454 \\
  $F_{\eta'\rightarrow\omega\gamma^*}(0)$   & $0.460\pm0.03$  & 0.450 & 0.470  & 0.461 \\

  $F_{\eta\rightarrow\gamma\gamma^*}(0)$  & $0.272\pm0.007$   & 0.272 & 0.259  & 0.277\\
  $F_{\eta'\rightarrow\gamma\gamma^*}(0)$ & $0.342\pm0.006$   & 0.342 & 0.317 & 0.334 \\
  $F_{\rho\rightarrow\eta\gamma^*}(0)$    & $1.59\pm0.05$     & 1.59 & 1.66  & 1.50 \\
  $F_{\eta'\rightarrow\rho\gamma^*}(0)$    & $1.35\pm0.06 $   & 1.35 & 1.42 & 1.39 \\

\hline
  $\theta^V$
& & $\begin{array}{c}
        \theta^V_8=12.17^\circ \\ \theta^V_0=77.82^\circ
         \end{array}$
& $\begin{array}{c}
        \theta^V_q=86.71^\circ \\ \theta^V_s=93.43^\circ
         \end{array}$
& $\begin{array}{c}
        \beta_V=-6.20^\circ \\
        \delta_V=1.40^\circ \\
        \gamma_V=-3.70^\circ
         \end{array}$\\
 \hline
 $\theta^S$
 &
 & $\begin{array}{c}
        \theta^S_8=-26.18^\circ \\ \theta^S_0=-2.85^\circ
   \end{array}$
& $\begin{array}{c}
        \theta^S_q=40.57^\circ \\ \theta^S_s=43.89^\circ
   \end{array}$
& $\begin{array}{c}
        \beta_S=-4.20^\circ \\
        \delta_S=-1.41^\circ \\
        \gamma_S=39.40^\circ
         \end{array}$\\
\hline
 $\begin{array}{c}
   \mathrm{Parameters} \\
   A(\mathrm{GeV}^{-1}) \\
   \beta (\mathrm{GeV})
 \end{array}$
 &
 & $\begin{array}{c}
    m_u=0.198\\
    m_s=0.556\\
    A_K=68.73 \\
     \beta_K=0.405\\
    A_\pi=47.36\\
    \beta_\pi=0.411\\
    A_{\eta8}=41.65\\
    \beta_{\eta8}=0.607\\
    A_{\eta0}=32.12 \\
    \beta_{\eta0}=0.925\\
     A_\rho=48.585\\
     \beta_\rho=0.373\\
    A_{\omega8}=215.18 \\
    \beta_{\omega8}=0.332\\
    A_{\omega0}=135.52\\
    \beta_{\eta0}=0.358
\end{array}$
&$\begin{array}{c}
     m_u=0.198\\
    m_s=0.556\\
    A_K=68.73 \\
     \beta_K=0.405\\
    A_\pi=47.36\\
    \beta_\pi=0.411\\
    A_{\eta q}=34.40\\
    \beta_{\eta q}=0.525\\
    A_{\eta s}=91.39\\
    \beta_{\eta s}=0.525\\
     A_\rho=48.585\\
     \beta_\rho=0.373\\
    A_{\omega q}=51.58\\
    \beta_{\omega q}=0.330\\
    A_{\omega s}=52.28\\
    \beta_{\omega s}=0.490
\end{array}$
&$\begin{array}{c}
    m_u=0.198 \\
    m_s=0.556 \\
    A_{K}=68.54\\
    \beta_{K}= 0.407\\
    A_{\pi_I}=47.36\\
    \beta_{\pi_I}=0.410\\
    A_{\eta q}=38.79\\
    \beta_{\eta q}= 0.486\\
    A_{\eta s}=95.46\\
    \beta_{\eta s}= 0.486\\
    A_{\rho_I}=38.14  \\
    \beta_{\rho_I}=0.411\\
    A_{\omega_I}=41.45\\
    \beta_{\omega_I}=0.419\\
    A_{\phi_I}=63.16\\
    \beta_{\phi_I}=0.476
\end{array}$\\
\hline $\chi^2 =
\sum(\frac{F_{\mathrm{th}}-F_{\mathrm{exp}}}{F_{\mathrm{exp}}})^2$ &
& 0.0786 & 0.0259 & 0.0083
\end{tabular}
\vspace*{5cm}  
\end{table*}

The $Q^2$ behavior of meson form factors is recalculated, as shown
in Figs.~\ref{fig1}-\ref{fig4}. In comparison with experimental
data, the $Q^2$ behavior fit data well in the low $Q^2$ region.
Extrapolating $Q^2$ to the time-like region by $q_\perp\rightarrow i
q_\perp $~\cite{Choi01}, we also recalculate the form factors in the
limited time-like region. The results are comparable with
experimental data, as shown in Fig.~\ref{fig5} and Fig.~\ref{fig6}.

In our calculation, only the relation got in the $Q^2\rightarrow
\infty$ limit is used to determine parameters, and perturbative QCD
effect is not actually taken into account when the form factors are
calculated. So, in fact the $Q^2 F_{P\gamma\gamma^*}(Q^2)$ and
$Q^2F_P(Q^2)$ calculated in our paper should only be valid when
$Q^2$ ranges from 0 to around a few GeV$^2$.

\begin{figure}
\resizebox{0.50\textwidth}{!}{%
\includegraphics[0,0][300,250]{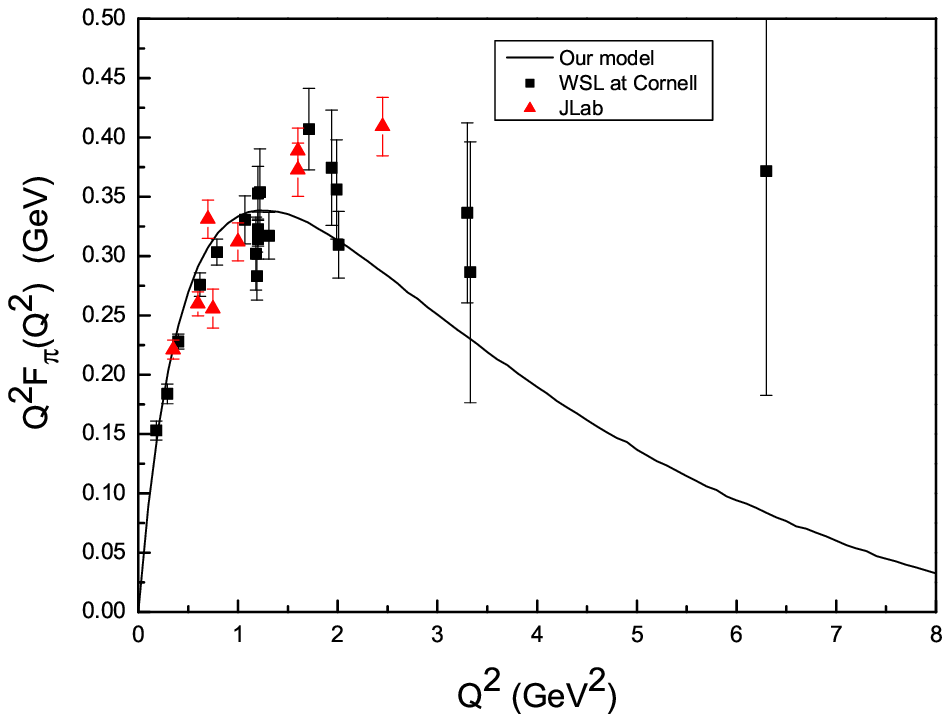}
}
\caption{\label{fig1} The $Q^2$ behavior of the form factor $Q^2
F_\pi(Q^2)$ compared with experimental data~\cite{Bebek78,Huber08}.
The solid squares are data measured by the Wilson Synchrotron
Laboratory at Cornell University. The solid triangles are data given
by JLab. }
\label{fig1}       
\end{figure}

\begin{figure}
\resizebox{0.50\textwidth}{!}{%
\includegraphics[0,0][300,250]{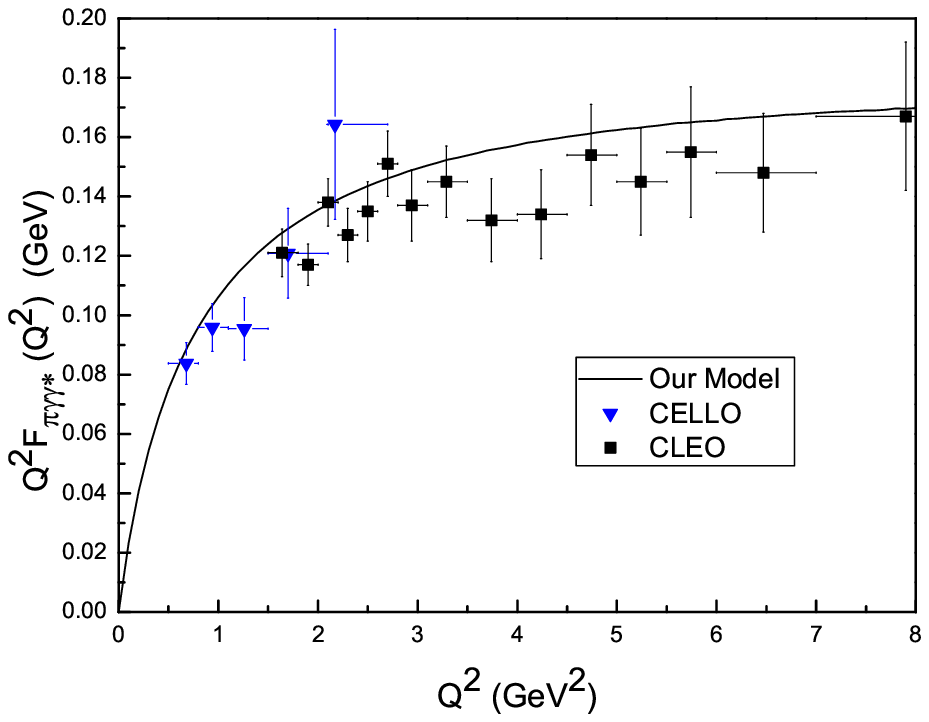}
} \caption{The $Q^2$ behavior of the form factor $Q^2
F_{\pi\rightarrow\gamma\gamma^*}(Q^2)$ compared with experimental
data~\cite{Behrend91,Gronberg98}.}
\label{fig2}       
\end{figure}

\begin{figure}
\resizebox{0.50\textwidth}{!}{%
\includegraphics[0,0][300,250]{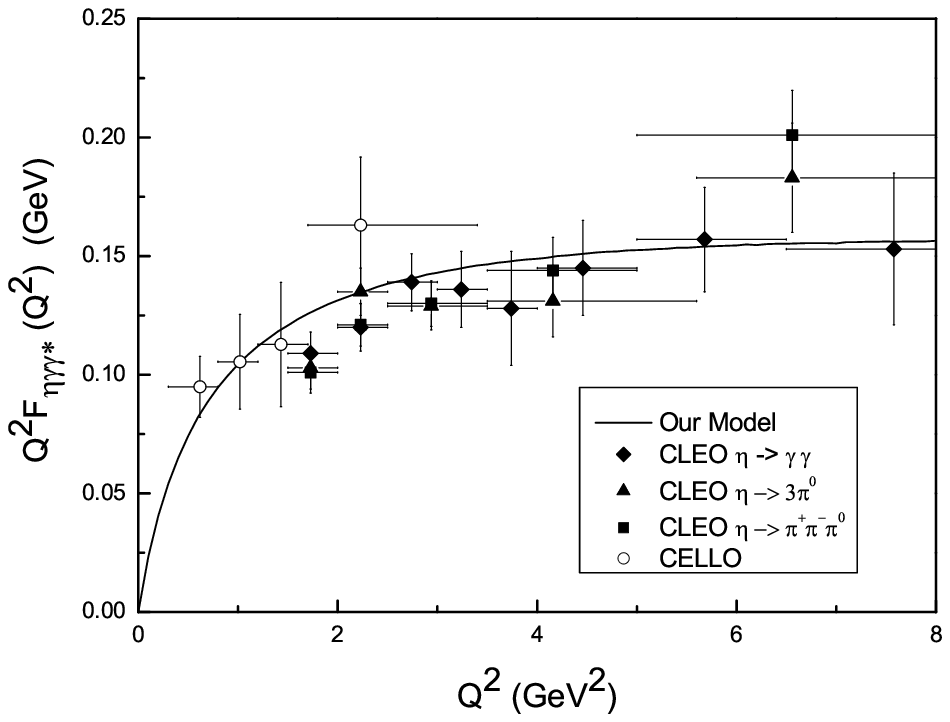}
}
\caption{The $Q^2$ behavior of the form factor $Q^2
F_{\eta\rightarrow\gamma\gamma^*}(Q^2)$ compared with experimental
data~\cite{Behrend91,Gronberg98}}
\label{fig3}       
\end{figure}

\begin{figure}
\resizebox{0.50\textwidth}{!}{%
\includegraphics[0,0][300,250]{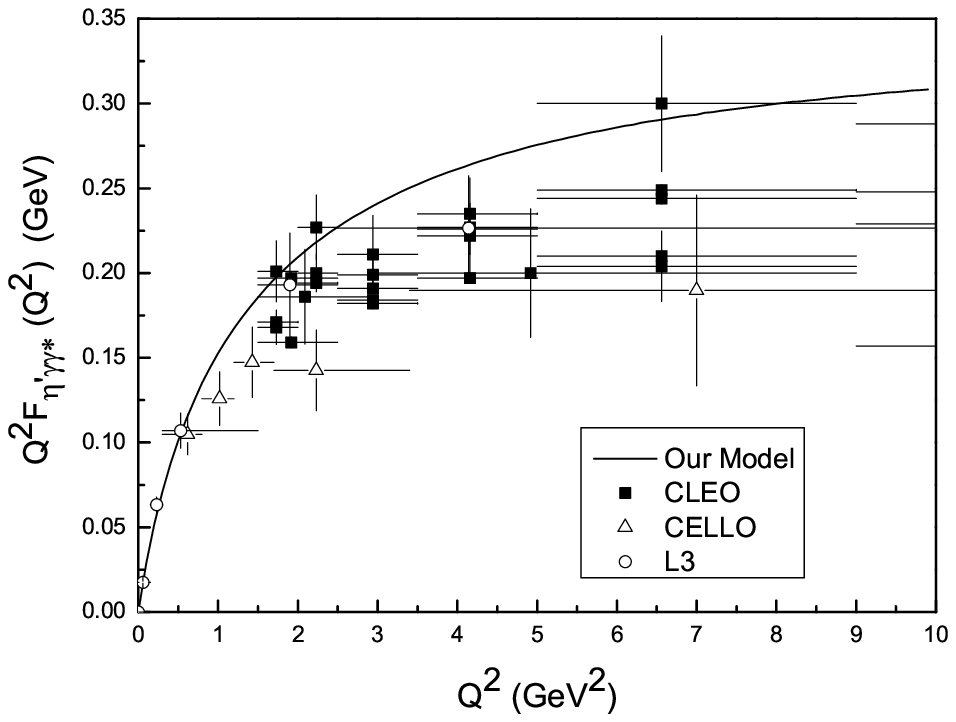}
} \caption{ The $Q^2$ behavior of the form factor $Q^2
F_{\eta'\rightarrow\gamma\gamma^*}(Q^2)$ compared with experimental
data~\cite{Behrend91,Gronberg98,Acciarri98}.}
\label{fig4}       
\end{figure}

\begin{figure}
\resizebox{0.45\textwidth}{!}{%
\includegraphics[0,0][300,250]{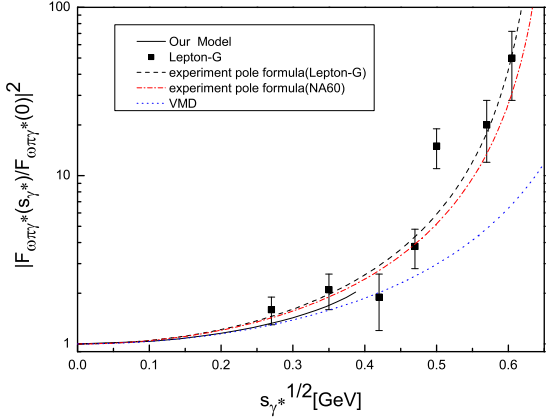}
} \caption{  The $Q^2$ behavior of the form factor $Q^2
F_{\omega\rightarrow\pi\gamma^*}(Q^2)$ compared with experimental
data. The dashed-dotted line is the result of fitting NA60 data with
the pole dependence~\cite{Arnaldia09} and the dashed line is
 that of fitting Lepton G data~\cite{Landsberg85}.
The dotted line is the vector meson dominance (VMD) model
prediction. }
\label{fig5}       
\end{figure}

\begin{figure}
\resizebox{0.50\textwidth}{!}{%
\includegraphics[0,0][300,250]{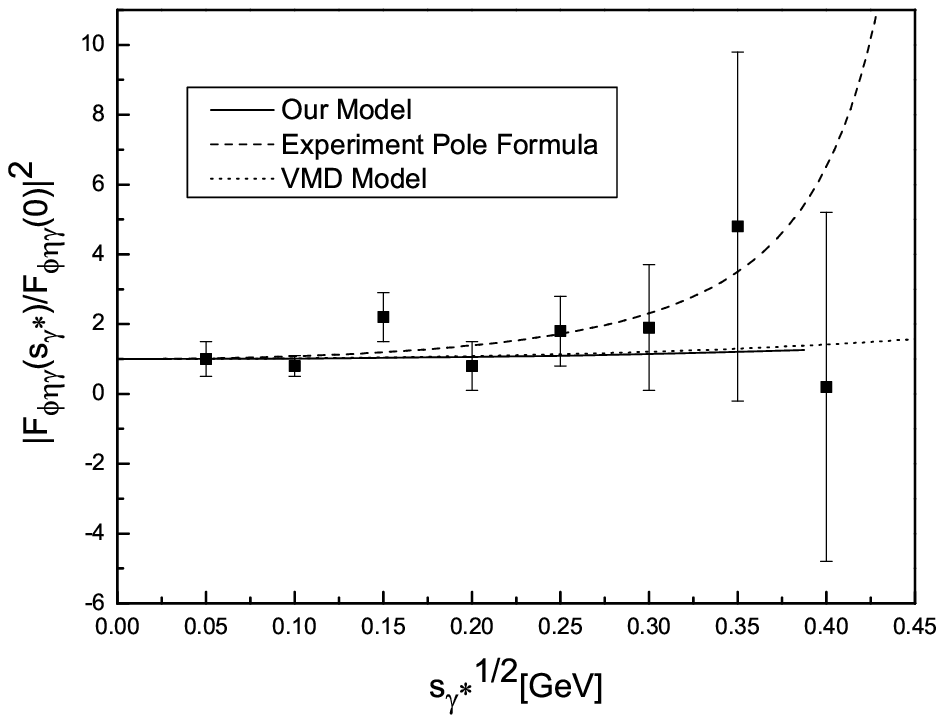}
} \caption{   The $Q^2$ behavior of the form factor $Q^2
F_{\phi\rightarrow\eta\gamma^*}(Q^2)$ compared with experimental
data~\cite{Achasov01}. The dashed-dotted line is the result of
fitting experimental data with the pole dependence. The dotted line
is the VMD model prediction. }
\label{fig6}       
\end{figure}

\section{\label{sec:level5}Conclusion}
In this paper we use the light-cone constituent quark model to
calculate the properties of the pseudoscalar mesons
$\pi$-$\eta$-$\eta'$ and the vector mesons $\rho$-$\omega$-$\phi$
with tri-meson-mixing pattern, which is described by unitary
tri-mixing matrices. All parameters including $m_u$, $m_s$, the wave
function parameters of $\pi$, $K$, $\eta_q$, $\eta_s$, $\rho_I$,
$\omega_I$, $\phi_I$ are constrained together by the experimental
data, including their electromagnetic decay widths and radii. The
reproduction of the experimental decay widths turn out to be better
than only two-meson mixing of $\eta$-$\eta'$ and $\omega$-$\phi$
were introduced, as can be seen from $\chi$'s in
Table~\ref{tab:table1}. We recalculate the $Q^2$ behavior of the
radiative transition form factors, and extrapolate them to the
limited time-like region. Results are comparable with experimental
data. The tri-meson-mixing pattern in this paper can explain why the
two-mixing-angle scenario with non-unitary mixing matrices of two
mesons could give acceptable results in our previous
work~\cite{Qian08}.

\section*{Acknowledgment}

This work is partially supported by National Natural Science
Foundation of China (No.~10721063 and No.~10975003) and by the Key
Grant Project of Chinese Ministry of Education (No.~305001).


\begin{thebibliography}{}

\bibitem{Gell-Mann64Zweig64}M. Gell-Mann, Phys. Lett. \textbf{8}, (1964) 214; G. Zweig, \textit{ An SU(3) Model} (1964).
\bibitem{Gell-Mann61}M. Gell-Mann, \textit{ The eightfold way}, W. A. Benjamin,
NY (1961).
\bibitem{Ne'eman61}Y. Ne'eman, Nucl. Phys. \textbf{26},(1961) 222.

\bibitem{Gell-Mann62Okubo62}M. Gell-Mann, Phys. Rev. 125, (1962) 1067;
S. Okubo, Progr. Theoret. Phys.  \textbf{27}, (1962) 949.

\bibitem{Sakurai62Coleman64}J. J. Sakurai, Phys. Rev. Lett. \textbf{9}, (1962) 472
; S. Coleman, Phys. Rev. \textbf{135}, (1964) B779.

\bibitem{Okubo63}S. Okubo, Phys. Lett. \textbf{5},(1963) 165;
J. Harte and G. Sachs, Phys. Rev. \textbf{135}, (1964) B459; L.
Epele, H. Fanchiotti and A. G. Grunfeld, Eur. Phys. J. C
\textbf{30},(2004) 97; A. Kucukarslan and Ulf-G. Meissner, Mod.
Phys. Lett. A \textbf{21}, (2006) 1423.

\bibitem{Castro96}G. Lopez Castro and  D.A. Lopez Falcon, Phys. Rev. D \textbf{54},(1996) 4400;
K.R. Nasriddinov, B.N. Kuranov, G.G. Takhtamyshev and T. A.
Merkulova, Phys. Atom. Nucl. \textbf{64}, (2001) 1326, Yad. Fiz.
\textbf{ 64}, (2001) 1402; M. Benayoun, P. David, L. Del Buono, O.
Leitner and H.B. O'Connell, Nucl. Phys. Proc. Suppl. \textbf{181},
(2008) 161.

\bibitem{Isgur76}N. Isgur, Phys. Rev. D \textbf{13}, (1976) 122.
\bibitem{Fritzsch77}H. Fritzsch and J.D. Jackson, Phys. Lett. B \textbf{66}, (1977) 365.
\bibitem{Kawarabayashi81}K. Kawarabayashi and N. Ohta, Nucl. Phys. B\textbf{175}, (1980) 477; Prog. Theor. Phys. \textbf{66}, (1981)
1789.

\bibitem{Leutwyler98}H. Leutwyler, Nucl. Phys. Proc. Suppl. \textbf{64},
(1998) 223.
\bibitem{Kaiser98}R. Kaiser and  H. Leutwyler, arXiv:hep-ph/9806336.
\bibitem{Feldmann98}T. Feldmann and P. Kroll, Eur. Phys. J. C \textbf{5}, (1998) 327.
For earlier treatments of two mixing angles for the $\eta$-$\eta'$
system, see, e.g., J. Schechter, A. Subbaraman, and H. Weigel, Phys.
Rev. D 48, (1993) 339.


\bibitem{Okubo77}S. Okubo, Phys. Rev. D \textbf{16}, (1977) 2336;
J. Jousset {\it et al.}, (DM2 Collaboration), Phys. Rev. D
\textbf{41}, (1990) 1389 ; N. Morisita, I. Kitamura and T. Teshima,
Phys. Rev. D \textbf{44}, (1991) 175; R. Escribano,
arXiv:hep-ph/0807.4201.

\bibitem{Glashow61}S. L. Glashow, Phys. Rev. Lett. \textbf{7}, (1961) 469 .

\bibitem{Coon87}S. A. Coon and R. C. Barrett, Phys. Rev. C \textbf{36}, (1987) 2189 ;
S. A. Coon, M. D. Scadron and P. C. McNamee, Nucl. Phys. A \textbf{
287}, (1977) 381; P. C. McNamee, M. D. Scadron and S. A. Coon, Nucl.
Phys. A \textbf{249},  (1975) 483; G. Krein, A. W. Thomas and A. G.
Williams, Phys. Lett. B \textbf{317}, (1993) 293; K. L. Mitchell,
P.C. Tandy, C. D. Roberts and R. T. Cahill, Phys. Lett. B
\textbf{335},(1994) 282 ; R. Machleidt and H. Muther, Phys. Rev. C
\textbf{63}, (2001) 034005 .


\bibitem{Enomoto}R. Enomoto and  M. Tanabashi, Phys. Lett. B \textbf{386}, (1996) 413.
\bibitem{Gardner98}S. Gardner, H. B. O'Connell and A. W. Thomas, Phys. Rev. Lett.
\textbf{80},  (1998) 1834.
\bibitem{Guo07}X.-H. Guo and  Z.-H. Zhang,  Phys. Rev. D\textbf{75},(2007) 074028 .
\bibitem{Guo00}X.-H. Guo and A. W. Thomas, Phys. Rev. D\textbf{61}, (2000) 116009 .

\bibitem{Maltman94} K. Maltman and T. Goldman, Nucl. Phys. A \textbf{572}, (1994) 682 .

\bibitem{Piekarewicz93}J. Piekarewicz, Phys. Rev. C \textbf{48},(1993) 1555
; T. Hatsuda, E.M. Henley, T. Meissner and G. Krein,  Phys. Rev. C
\textbf{49}, (1994) 452 ; M.J. Iqbal and J.A. Niskanen, Phys. Lett.
B \textbf{322},(1994) 7 ; H. B. O'Connell, B.C. Pearce, A. W. Thomas
and A. G. Williams, Phys. Lett. B\textbf{336}, (1994)1 ; A. N. Mitra
and K.C. Yang, Phys. Rev. C\textbf{51}, (1995) 3404 ; T. D. Cohen
and G. A. Miller, Phys. Rev. C\textbf{52}, (1995) 3428 .



\bibitem{Benayoun01}M. Benayoun and H. B. O'Connell, Eur. Phys. J. C
\textbf{22}, (2001) 503.
\bibitem{Benayoun08b}M. Benayoun, P. David, L. DelBuono, O. Leitner and H. B.
O'Connell, Eur. Phys. J. C \textbf{55}, (2008) 199.
\bibitem{Benayoun08c}M. Benayoun, arXiv:hep-ph/0805.1835.
\bibitem{Gusbin81}D. Gusbin, Phys. Rev. D \textbf{24}, (1981)797.

\bibitem{Lepage80}G. P. Lepage and S. J. Brodsky, Phys. Rev. D \textbf{22}, (1980) 2157.
\bibitem{Brodsky82}S. J. Brodsky, T. Huang, G. P. Lepage, in \textit{ Quarks and Nuclear Forces},
        edited by D. Fries and B. Zeitnitz (Springer, Tracts in Modern Physics, Vol. 100) (Springer, New York, 1982);
        S. J. Brodsky, T. Huang and G. P. Lepage, in Particles and Fields-2,
             edited by A. Z. Capri and A. N. Kamal (Plenum, New York, 1983), p.143.
\bibitem{Brodsky98}S. J. Brodsky, H.-C. Pauli and S. S. Pinsky, Phys. Rep.
\textbf{301}, (1998) 299.

\bibitem{Choi97a}H.-M. Choi and C.-R. Ji, Phys. Rev. D \textbf{56}, (1997) 6010.
\bibitem{Cao98}J. Cao, F.-G. Cao, T. Huang and B.-Q. Ma, Phys. Rev. D\textbf{58}, (1998) 113006.
\bibitem{Choi99}H.-M. Choi and C.-R. Ji, Phys. Rev. D\textbf{59}, (1999) 074015.
\bibitem{Xiao05}B.-W. Xiao and B.-Q. Ma, Phys. Rev. D \textbf{68}, 034020 (2003); Phys. Rev. D\textbf{71}, (2005) 014034.
\bibitem{Huang07}T. Huang and X.-G. Wu, Eur. Phys. J. C \textbf{50}, (2007) 771.


\bibitem{Melosh74}H. J. Melosh, Phys. Rev. D \textbf{9}, (1974) 1095; E. Wigner, Ann. Math. \textbf{40}, (1939) 149.
\bibitem{Kondratyuk80}L.~A.~Kondratyuk and M.~V.~Terentev, Sov. J. Nucl. Phys.  \textbf{31},  (1980) 561
                     [Yad. Fiz. \textbf{31}, (1980) 1087].
\bibitem{Ma93}B.-Q. Ma, J. Phys.  G  \textbf{17}, (1991)  L53 [arXiv:hep-ph/0711.2335]; B.-Q.
Ma, Q.-R. Zhang, Z. Phys. C \textbf{ 58}, (1993) 479; B.-Q. Ma, Z.
Phys. A \textbf{345}, (1993) 321.

\bibitem{Donoghue85}J. F. Donoghue, B. R. Holstein and Y. C. R. Lin, Phys. Rev. Lett. \textbf{55}, (1985) 2766.
\bibitem{Gilman87}F. J. Gilman and R. Kauffman, Phys. Rev. D \textbf{36}, (1987) 2761.

\bibitem{Feldmann99}Th. Feldmann, P. Kroll and B. Stech, Phys. Lett. B \textbf{449}, (1999) 339.
\bibitem{Cao99}F.-G. Cao and A. I. Signal, Phys. Rev. D \textbf{60},  (1999) 114012.
\bibitem{Qian08}W. Qian and B. -Q. Ma, Phys. Rev. D \textbf{78}, (2008) 074002.
\bibitem{Benayoun99}M. Benayoun, L. DelBuono, S. Eidelman, V. N.
Ivanchenko and H. B. O'Connell, Phys. Rev. D \textbf{59}, (1999)
114027.
\bibitem{Huber08}G. M. Huber {\it et al.}, (Jefferson Lab $F_\pi$
Collaboration), Phys. Rev. C \textbf{78},  (2008) 045203.
\bibitem{Bebek78}C. J. Bebek, C. N. Brown and S. D. Holmes {\it et
al.}, Phys. Rev. D \textbf{17}, (1978) 1693.
\bibitem{Behrend91}H.-J. Behrend {\it et al.}, (CELLO Collaboration), Z. Phys. C \textbf{49}, (1991) 401.
\bibitem{Gronberg98}J. Gronberg {\it et al.}, (CLEO Collaboration),
Phys. Rev. D \textbf{57}, (1998) 33.
\bibitem{Acciarri98}M. Acciarri {\it et al.}, (L3 Collaboration),
Phys. Lett. B \textbf{418}, (1998) 399.
\bibitem{Feldmann00}T. Feldmann, Int. J. Mod. Phy. A \textbf{15},
(2000) 159.
\bibitem{Escribano05}R. Escribano, and  J. Frere, JHEP \textbf{06},
(2005) 029.
\bibitem{Huang94}T. Huang, B.-Q. Ma and Q.-X. Shen, Phys. Rev. D \textbf{49}, (1994) 1490.

\bibitem{Jaus91}W. Jaus, Phys. Rev. D \textbf{44}, (1991) 2851.
\bibitem{Choi97}H.-M. Choi and C.-R. Ji, Nucl. Phy. A \textbf{618}, (1997) 291.
\bibitem{PDG08}C. Amsler, {\it et al.} (Particle Data Group), Phys. Lett. B \textbf{667}, (2008) 1.
\bibitem{Yu07}J.-H. Yu, B.-W. Xiao and B.-Q. Ma, J. Phys. G \textbf{34}, (2007) 1845.
\bibitem{Choi01}H.-M. Choi and C.-R. Ji, Nucl. Phys. A \textbf{679},  (2001) 735.
\bibitem{Arnaldia09}R. Arnaldi {\it et al.} (NA60 Collaboration), arXiv:hep-ph/0902.2547.
\bibitem{Landsberg85}L.G. Landsberg, Phys. Rep. \textbf{128}, (1985) 301.
\bibitem{Achasov01}A. N. Achasov {\it et al.}, Phys. Lett. B \textbf{
504}, (2001) 275.
\end{thebibliography}
\end{document}